\documentclass[a4paper,fleqn,usenatbib]{mnras}

\usepackage{newtxtext,newtxmath}

\usepackage[T1]{fontenc}
\usepackage{ae,aecompl}

\usepackage{graphicx}	
\usepackage{amsmath}	
\usepackage{amssymb}	

\defcitealias{Diemer2014}{DK14}
\defcitealias{Hoekstra2015}{H15}

\title[Weak lensing constraints on splashback]{Weak lensing constraints on splashback around massive clusters}

\author[O. Contigiani et al.]{
Omar Contigiani$^{1,2}$,\thanks{E-mail: contigiani@strw.leidenuniv.nl} 
Henk Hoekstra$^{1}$,
Yannick M. Bah\'e$^{1}$
\\
$^{1}$Leiden Observatory, Leiden University, PO Box 9513, 2300 RA, Leiden, the Netherlands
\\
$^{2}$Lorentz Institute for Theoretical Physics, Leiden University, PO Box 9506, 2300 RA, Leiden, The Netherlands
}

\date{Accepted XXX. Received YYY; in original form ZZZ}

\pubyear{2018}

\begin{document}
\label{firstpage}
\pagerange{\pageref{firstpage}--\pageref{lastpage}}
\maketitle

\begin{abstract}
The splashback radius $r_\text{sp}$ separates the physical regimes of collapsed and infalling material around massive dark matter haloes. In cosmological simulations, this location is associated with a steepening of the spherically averaged density profile $\rho(r)$.
In this work, we measure the splashback feature in the stacked weak gravitational lensing signal of $27$ massive clusters from the Cluster Canadian Comparison Project with careful control of residual systematics effects.
We find that the shear introduced by the presence of additional structure along the line of sight significantly affects the noise at large clustercentric distances. 
Although we do not  detect a significant steepening, the use of a simple parametric model enables us to measure both $r_\text{sp}=3.5^{+1.1}_{-0.7}$ comoving Mpc and the value of the logarithmic slope $\gamma = \log \rho / \log r$  at this point, $\gamma(r_\text{sp}) = -4.3^{+1.0}_{-1.5}$. 
\end{abstract}




\section{Introduction}
\label{sec:Intro}
In the concordance $\Lambda$CDM model, collisionless dark matter acts as the building block of cosmic structure, contributing about $25\%$ of the total energy density in the Universe and the majority of the total mass \citep[Planck 2015 results,][]{Ade2016}. In this framework, gravity is the primary force behind the growth of structure in the matter field and is able to form the present-day cosmic web from an almost homogeneous initial state. Fully collapsed structures, known as haloes, are thought to grow both through mergers of smaller ones (hierarchical clustering) and continuous infall of ambient dark matter (smooth accretion). 

An intuitive understanding of this second mechanism is given by the study of spherical collapse in an expanding Universe  \citep[see ][for some historic landmark results]{Gunn1972, Fillmore1984}. Shells of material surrounding an overdensity eventually decouple from the Hubble flow and start collapsing towards it. As more shells orbit the halo, the wrapping in phase-space of different streams results in caustics visible in the density profile. Of particular interest is the region around the outermost caustic, where the physical regimes of accreting and collapsed material meet. 

More recently, \citet[DK14 from now on]{Diemer2014} studied the spherically averaged density profile $\rho(r)$ of these regions in dark matter only simulations and have reported a change in slope compared to the collisionless equilibrium profile \citep[Einasto or NFW,][]{Einasto1965, Navarro1997}. \cite{More} argued that the splashback radius $r_\text{sp}$, corresponding to the minimum logarithmic slope $\gamma(r) = \log \rho(r) / \log r$, could function as a physically motivated definition for the boundary of dark matter haloes. This role is usually assumed by proxies for the virial radius such as $r_{200m}$, defined as the radius inside which the average halo density is $200$ times the average matter density of the Universe $\rho_m$. While this radius has a clear definition based on analytical solutions of idealized virialization scenarios, the mass contained within it, known as $M_{200m}$, is an imperfect measure of the halo mass. This is because it is subject to a pseudo-evolution caused by the redshift dependence of $\rho_m$ \citep{Diemer2013}. In contrast, because the caustic associated with splashback is connected to the apocenter of recently accreted material, all the material within $r_\text{sp}$ is necessarily collapsed material and should rightfully contribute to the halo mass.

At larger distances, the presence of correlated structure surrounding the halo is expected to shape the density profile. Using the language of the halo model \citep[see, e.g.,][ for a review]{Cooray2002}, this is a transition region from the 1-halo term to the 2-halo term. \citetalias{Diemer2014} have however reported that in the outermost regions ($r \lesssim 9 r_{200m}$), the 2-halo term based on the matter correlation function provides a worse fit to simulations compared to a simple power-law.  

Because the slope of the density profile at $r_\text{sp}$ is found to be, on average, a decreasing function of the halo mass, \citetalias{Diemer2014} first pointed out that large overdensities are the ideal target for the detection of this feature -- i.e., measuring a significant departure from the equilibrium profile. This makes galaxy clusters the ideal candidates since they correspond to the most massive haloes in the Universe. For this mass range, $r_\text{sp}$ is expected to be located around $r_{200m}$, at a cluster-centric distance of the order of a few Mpc. 

The splashback feature should also be present in the radial distribution of galaxies. This was first detected by \cite{More2016} using the large sample of redMaPPer clusters from \cite{Rykoff2014}, and studied further in 
\cite{Baxter2017}. However, these studies find a discrepancy between the inferred splashback radius and the expected distribution of subhaloes based on dark matter only simulations. Known physical processes (e.g., tidal disruption and dynamical friction) are not expected to induce a mismatch between the galaxy and subhalo distributions at splashback scales and this deviation is still unexplained. In particular, while the results have been shown to depend on the details of the cluster finding algorithm \citep{Zu2017, Busch2017}, it is still uncertain if this can fully explain the discrepancy \citep{Chang2018}.

\cite{Chang2018} studied a sample of redMaPPer clusters selected in Dark Energy Survey year 1 data. For this large sample, they detected a splashback feature in the galaxy distribution {\it and} from weak lensing measurements. The latter has the advantage that the lensing signal probes the matter distribution directly \citep[see e.g.][for a review]{Hoekstra2013a}. The first attempt to detect the splashback feature using weak gravitational lensing was presented in \citet{Umetsu2016}, who used a sample of $16$ high-mass clusters in the Cluster Lensing and Supernova survey with Hubble (CLASH). Unfortunately, the limited field of view (foV) of Suprime-Cam prevented precise measurements in the outer regions, and as a result, \citet{Umetsu2016} could only provide a lower limit on the splashback radius. 

In this work, we provide a measurement\footnote{In the interest of reproducibility we make our splashback code publicly available at \url{https://github.com/contigiani/splash/}}  of splashback using weak lensing observations for a sample of 27 massive clusters of galaxies that were observed as part of the Canadian Cluster Comparison Project \citep[CCCP;][]{Hoekstra2012}. Hence our strategy is similar to that employed by \cite{Umetsu2016}, but we take advantage of the fact that the CCCP observations were obtained using 
MegaCam, which has a foV of 1~deg$^2$, and enables us to measure the lensing signal beyond the splashback radius. The paper is organized as follows: in Section~\ref{sec:Data} we present our dataset and describe our lensing analysis, in Section~\ref{sec:splash} we show the results of our fit and the implications for splashback, and in Section~\ref{sec:Conclusions} we draw our conclusions.  Throughout the paper we employ a flat $\Lambda$CDM cosmology with $H_0 = 70~\text{Mpc/km/s}$, $\Omega_m = 0.3$, $\Omega_{c}=0.25$ and $\sigma_8 =0.80$.

\section{Cluster lensing}
\label{sec:Data}

In this section, we discuss how the sheared images of distant galaxies can be used to constrain the matter distribution of clusters along the line of sight. After introducing our cluster sample, we present the weak lensing measurements and explain our methodology, with a particular focus on systematic effects and noise estimation. 

\subsection{Sample characterization}

\begin{table*}
	\begin{tabular}{l|l|l|c|c|c|c}
		\hline
			Name & RA & DEC & $z$ & $\langle \beta \rangle$ & $M_\text{g}$ & M$_\text{200m}$ \\
           		& (J2000) & (J2000) & & & [$10^{13}$ M$_\odot$] & [$10^{14}$ M$_\odot$] \\
		\hline
MS 0440.5+0204 & $04^\mathrm{h}43^\mathrm{m}09.0^\mathrm{s}$ & $+02$\degr $10$\arcmin $19$\arcsec & 0.19 & 0.656 & 2.4 & 3.8 \\
Abell 1234 & $11^\mathrm{h}22^\mathrm{m}30.0^\mathrm{s}$ & $+21$\degr $24$\arcmin $22$\arcsec & 0.163 & 0.699 & 3.8$^\dagger$ & 8.3 \\
RX J1524.6+0957 & $15^\mathrm{h}24^\mathrm{m}38.3^\mathrm{s}$ & $+09$\degr $57$\arcmin $43$\arcsec & 0.516 & 0.329 & 4.1 & 6.5 \\
Abell 1942 & $14^\mathrm{h}38^\mathrm{m}21.9^\mathrm{s}$ & $+03$\degr $40$\arcmin $13$\arcsec & 0.224 & 0.621 & 4.4 & 14.6 \\
Abell 2259 & $17^\mathrm{h}20^\mathrm{m}09.7^\mathrm{s}$ & $+27$\degr $40$\arcmin $08$\arcsec & 0.164 & 0.697 & 5.0 & 8.6 \\
MACS J0913.7+4056 & $09^\mathrm{h}13^\mathrm{m}45.5^\mathrm{s}$ & $+40$\degr $56$\arcmin $29$\arcsec & 0.442 & 0.396 & 5.3 & 6.8 \\
Abell 1246 & $11^\mathrm{h}23^\mathrm{m}58.5^\mathrm{s}$ & $+21$\degr $28$\arcmin $50$\arcsec & 0.19 & 0.661 & 5.6$^\dagger$ & 9.5\\
MS 1008.1-1224 & $10^\mathrm{h}10^\mathrm{m}32.3^\mathrm{s}$ & $-12$\degr $39$\arcmin $53$\arcsec & 0.301 & 0.526 & 5.8 & 17.4 \\
3C295 & $14^\mathrm{h}11^\mathrm{m}20.6^\mathrm{s}$ & $+52$\degr $12$\arcmin $10$\arcsec & 0.46 & 0.374 & 6.2 & 12.6 \\
Abell 586 & $07^\mathrm{h}32^\mathrm{m}20.3^\mathrm{s}$ & $+31$\degr $38$\arcmin $01$\arcsec & 0.171 & 0.674 & 6.5 & 5.0 \\
Abell 611 & $08^\mathrm{h}00^\mathrm{m}56.8^\mathrm{s}$ & $+36$\degr $03$\arcmin $24$\arcsec & 0.288 & 0.533 & 6.6 & 10.0 \\
Abell 2104 & $15^\mathrm{h}40^\mathrm{m}07.9^\mathrm{s}$ & $-03$\degr $18$\arcmin $16$\arcsec & 0.153 & 0.712 & 6.8 & 17.2 \\
Abell 2111 & $15^\mathrm{h}39^\mathrm{m}40.5^\mathrm{s}$ & $+34$\degr $25$\arcmin $40.5$\arcsec & 0.229 & 0.614 & 7.4 & 10.2 \\
Abell 959 & $10^\mathrm{h}17^\mathrm{m}36.0^\mathrm{s}$ & $+59$\degr $34$\arcmin $02$\arcsec & 0.286 & 0.549 & 7.5 & 21.1 \\
\hline
Abell 520 & $04^\mathrm{h}54^\mathrm{m}10.1^{\mathrm{s}\S}$ & $+02$\degr $55$\arcmin $18$\arcsec$^{\S}$ & 0.199 & 0.654 & 8.5 & 16.6 \\
Abell 2537 & $23^\mathrm{h}08^\mathrm{m}22.2^\mathrm{s}$ & $-02$\degr $11$\arcmin $32$\arcsec & 0.295 & 0.532 & 8.6 & 22.4 \\
Abell 851 & $09^\mathrm{h}42^\mathrm{m}57.5^{\mathrm{s}\S}$ & $+46$\degr $58$\arcmin $50$\arcsec$^{\S}$ & 0.407 & 0.421 & 9.7 & 22.6 \\
Abell 1914 & $14^\mathrm{h}26^\mathrm{m}02.8^{\mathrm{s}\S}$ & $+37$\degr $49$\arcmin $28$\arcsec$^{\S}$ & 0.171 & 0.693 & 9.9 & 14.7 \\
MS 0451.6-0305& $04^\mathrm{h}54^\mathrm{m}10.8^\mathrm{s}$ & $-03$\degr $00$\arcmin $51$\arcsec & 0.54 & 0.315 & 10.3 & 18.0 \\
Abell 521 & $04^\mathrm{h}54^\mathrm{m}06.9^\mathrm{s}$ & $-10$\degr $13$\arcmin $25$\arcsec & 0.253 & 0.577 & 10.6 & 11.5 \\
Abell 2204 & $16^\mathrm{h}32^\mathrm{m}47.0^\mathrm{s}$ & $+05$\degr $34$\arcmin $33$\arcsec & 0.152 & 0.714 & 11.6 & 21.8 \\
Abell 1835 & $14^\mathrm{h}01^\mathrm{m}02.1^\mathrm{s}$ & $+02$\degr $52$\arcmin $43$\arcsec & 0.253 & 0.58 & 12.1 & 21.5 \\
Abell 2261 & $17^\mathrm{h}22^\mathrm{m}27.2^\mathrm{s}$ & $+32$\degr $07$\arcmin $58$\arcsec & 0.224 & 0.621 & 14.6 & 26.4 \\
CIZA J1938+54 & $19^\mathrm{h}38^\mathrm{m}18.1^\mathrm{s}$ & $+54$\degr $09$\arcmin $40$\arcsec & 0.26 & 0.569 & 15.6$^\dagger$ & 18.6 \\
Abell 697 & $08^\mathrm{h}42^\mathrm{m}57.6^\mathrm{s}$ & $+36$\degr $21$\arcmin $59$\arcsec & 0.282 & 0.552 & 15.6 & 15.1 \\
RX J1347.5-1145 & $13^\mathrm{h}47^\mathrm{m}30.1^\mathrm{s}$ & $-11$\degr $45$\arcmin $09$\arcsec & 0.451 & 0.377 & 16.3 & 20.9 \\
Abell 2163 & $16^\mathrm{h}15^\mathrm{m}49.0^\mathrm{s}$ & $-06$\degr $08$\arcmin $41$\arcsec & 0.203 & 0.63 & 23.3 & 18.9 \\
	\end{tabular}
 
\caption{The full cluster sample, ``CCCP all'', used in this paper. RA and DEC are the sky position of the cluster centre (brightest cluster galaxy, or X-ray peak for coordinates marked with $\S$), $z$ is the cluster redshift, $\langle\beta\rangle$ is the average value of $D_\text{LS}/D_\text{S}$ (see Sec.~\ref{sec:shear}), $M_{g}$ is the gas mass within $r_{500c}$, defined as the radius of the sphere inside which the mean halo density is $500$ times the critical density of the Universe at redshift $z$ and $M_{200m}$ is the mass enclosed within $r_{200m}$. These values are recovered from the NFW fit performed in H15. The values for $M_{g}$ are taken from the X-ray analysis of M13 or, for values marked with $\dagger$, they are defined using the scaling relations found in the same paper. Clusters listed below the horizontal line belong to the high mass subsample.}
\label{tab:sample}
\end{table*}

Our dataset is based on the Canadian Cluster Comparison Project (CCCP), a survey targeting X-ray selected massive clusters at $z\lesssim0.5$ introduced for the first time in \citet{Hoekstra2012} and re-analysed in \citet[H15 from now on]{Hoekstra2015}. The starting points of our analysis are the $r$-band images of $27$ clusters captured by MegaCam at the Canada-France-Hawaii Telescope (CHFT). We exclude from the original CCCP images those corresponding to on-going mergers: Abell 115, Abell 222/3, Abell 1758, and MACS J0717.5+3745 because these systems display a visible double peaked matter distribution for which two splashback surfaces might intersect each other. 

The objects are characterized by masses $3.8<M_{200m}/(10^{14}~\mathrm{M}_\odot)<26.4$ and cover a redshift range $0.15<z<0.55$, with only $6$ of them located at $z>0.3$. Table~\ref{tab:sample} reviews the sample and presents the quantities relevant for the present work. For more details about the cluster sample we refer the reader to \cite{Hoekstra2012}, \citetalias{Hoekstra2015} for a description of the weak lensing analysis, and the companion paper \cite{Mahdavi2013} for the analysis of X-ray observations.

In simulations, \citetalias{Diemer2014} found a correlation between the splashback feature and the halo mass. We, therefore, define a high-mass subsample of our clusters, containing the $13$ most massive objects. The average $M_{200m}$ of the sample and the subsample, weighted by the signal-to-noise ratio (SNR), equal $1.7$ and $2.0\times 10^{15}~\mathrm{M}_\odot$, respectively. We choose to employ the gas mass $M_g$ within $r_{500c}$ reported by \cite{Mahdavi2013} to define our high-mass threshold. This is because this value is found to be a robust estimator of the weak lensing mass and its measurement is mostly independent from it since it is based on a different physical mechanism. A weak dependence between the two is left due to the lensing-based definition of $r_{500c}$.

Targeted observations such as the ones discussed in this work currently represent the most efficient approach to study clusters of virial mass around $10^{15}$  M$_\odot$. In particular, such a sample cannot be obtained by present-day or near-future wide surveys, e.g., DES \citep{DESCollaboration2017} or the Kilo-Degree Survey \citep{DeJong2017}, because massive haloes are rare (i.e. $\ll 1$ per FoV) and targeted deep data result in a higher SNR compared to wide surveys. For these reasons, the SDSS and DES studies of \cite{More2016}, \cite{Baxter2017} and \cite{Chang2018} are based instead on large samples of low-mass clusters: $8649$ clusters with $\langle M_{200m}\rangle = 2.7\times 10^{14}$ M$_\odot$ for SDSS \citep{Miyatake2016} and $3684$ clusters with  $\langle M_{200m}\rangle=3.6\times 10^{14}$ M$_\odot$ for DES Y1. In contrast, our dataset is much closer in nature to the CLASH sample used in \cite{Umetsu2016}, also based on targeted observations. In particular, the mass of their stacked ensemble, $ M_{200m} = 1.9\times 10^{15}$ M$_\odot$, matches ours. Nevertheless, we want to mention one feature unique to CCCP: the FoV of MegaCam ($1\times1\deg$) is significantly larger than that of Suprime-Cam ($34\times27$ arcmin), the instrument used for the CLASH profile reconstruction at large scales \citep{Umetsu2016a}. This is particularly suited for our purposes since it allows us to better cover cluster-centric distances where the splashback radius is located.

\subsection{Tangential shear}
\label{sec:shear}

\begin{figure}
	\includegraphics[width=0.47\textwidth]{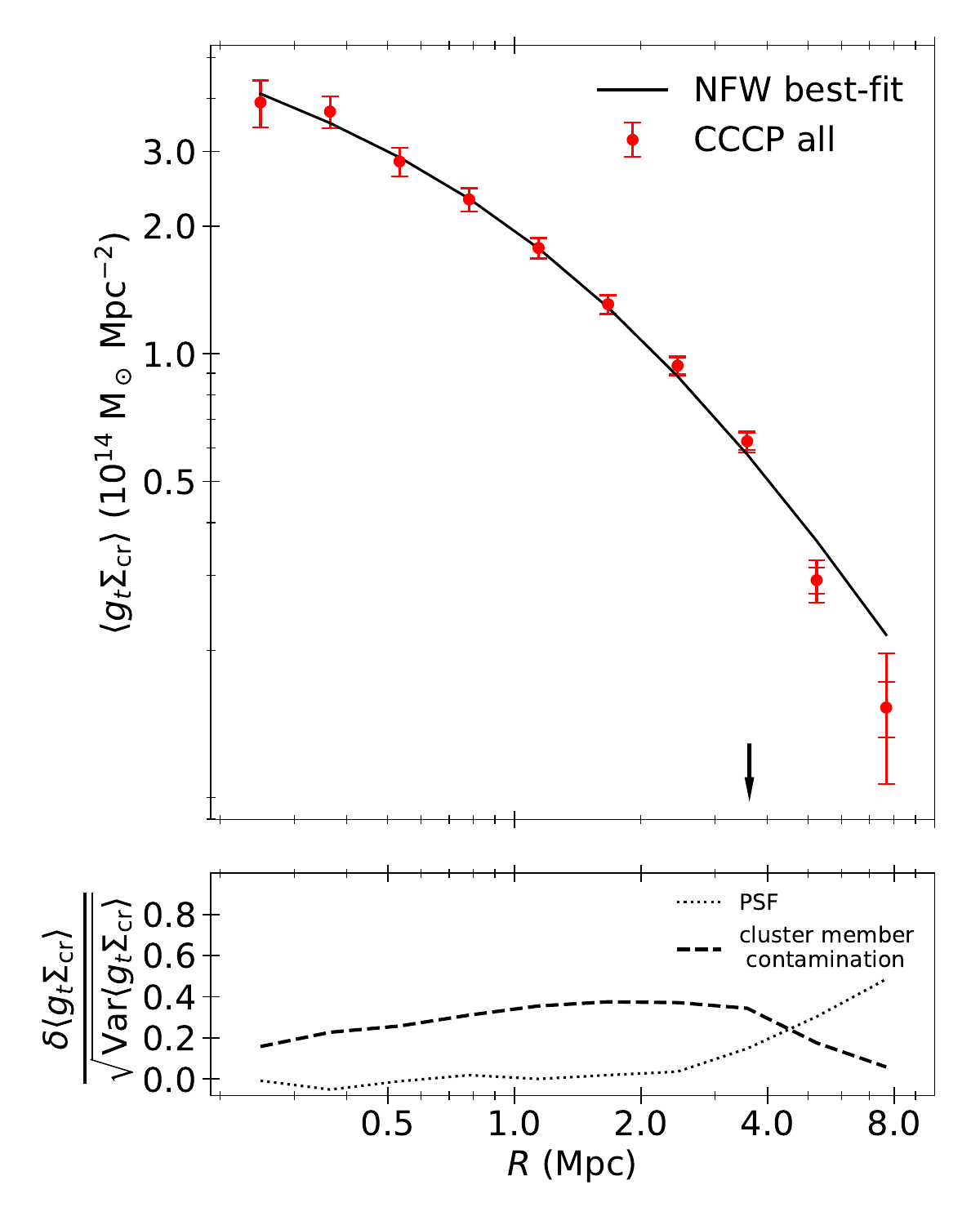}
	\caption{Lensing signal. The top panel shows the noise-weighted stacked signal of the $27$ clusters in our sample as a function of comoving clustercentric distance, together with a best fit NFW profile to the first five data points (see Sec~\ref{sec:shear} for more information). The arrow points to the inferred location of $r_{200m}$; in simulated galaxy clusters the splashback feature is located around this position. The larger error bars are the full $1\sigma$ errors for the data points, while the inner error bars account only for statistical uncertainty. The difference between the two is apparent only in the last few data points. The bottom panel shows an estimate of the expected residual systematics left after the corrections discussed in Sec.~\ref{sec:sys} are applied, expressed as a fraction of the total uncertainty. These effects are found to be consistent with the error bars.}
	\label{fig:shear}
\end{figure}

In the weak lensing regime, the shear field is found by averaging the PSF-corrected ellipticities of a sample of background sources. We follow \citetalias{Hoekstra2015} and use sources in the magnitude range $22<m_r<25$. The lower limit reduces the presence of foreground objects such as bright galaxies belonging to the clusters, which are abundant in the central regions and are not sheared by the cluster's mass distribution. Because this operation is unable to completely remove cluster members, we chose to model the residual contamination statistically, as explained in Sec.~\ref{sec:sys}. 

Shapes are measured using an improved KSB method \citep{Kaiser1995, Luppino1997, Hoekstra1998}. The quadrupole moments of the galaxy images are used to construct a polarization tensor $\mathbfit{e}$, which is then corrected for the  point spread function (PSF) of the observing instrument. In Section \ref{sec:sys} we address this step in more detail and mention the improvements we have implemented since \citetalias{Hoekstra2015}. The shear polarizability $\tilde{P}^\gamma$ quantifies how the observed polarization of an individual galaxy is related to the 
gravitational shear. For an ensemble of sources the shear components are hence measured as a noise-weighted average, $\langle e_i /\tilde{P}^\gamma \rangle $, where the individual weights are written as \citep{Hoekstra2000}
\begin{equation}
	w = \frac{1}{\langle \epsilon^2 \rangle  + \left( \sigma_e / \tilde{P}^\gamma\right)^2}.
\end{equation}
In this expression two sources of noise are included: the scatter introduced by the intrinsic variance of galaxy ellipticities $\langle \epsilon^2 \rangle$ and the uncertainty in the measured polarization $\sigma_e$ due to noise in the imaging data. Following \cite{Hoekstra2000} we use $\langle \epsilon^2 \rangle^{1/2} = 0.25$.

For an isolated circular overdensity, the induced shear is purely tangential, i.e., the deformation is parallel to the radial direction. In general, this shear component is related to the projected mass surface density $\Sigma(R)$ as a function of the radial coordinate $R$:
\begin{align}
&\gamma_{\text{t}}(R) =  \frac{\overline{\Sigma}(< R) - \Sigma(R)}{\Sigma_{\text{cr}}}  =  \frac{\Delta \Sigma(R)}{\Sigma_{\text{cr}}},
\label{eq:shear}
\\
&\Sigma_{\text{cr}} = \frac{c^2}{4\pi G}\frac{1}{\langle\beta\rangle} \frac{1}{D_{\text{L}}}.
\end{align}
In these expressions, the profile $\Delta \Sigma(R)$ is called excess surface density (ESD) and the critical density $\Sigma_{\text{cr}}$ is a geometrical factor quantifying the lensing efficiency as a function of the relative position of source and lens. The definition above applies for a lens at distance $D_\mathrm{L}$ shearing an ensemble of sources. $\langle\beta\rangle$ is the average of the quantity $\text{max}\left[0, D_\text{LS}/D_\text{S}\right]$ for each source, with $D_\text{LS}$ being the individual lens-to-source distance\footnote{Note that $D_\text{LS}$ is negative for foreground sources.} and $D_\text{S}$ the distance to the source.

Because we work with single-band observations, we are unable to derive individual photometric redshifts. Fortunately, a representative photometric redshift distribution is sufficient to estimate $\beta$. This distribution is obtained for all clusters by magnitude-matching the most recent COSMOS photometric catalogue \citep[COSMOS2015,][]{Laigle2016} to our source $r$-band magnitude range. 

We point out that the measured average ellipticity is an estimator of the reduced shear 
\begin{equation}
g_i = \frac{\gamma_i(R)}{1-\Sigma(R)/\Sigma_\text{crit}}.
\label{eq:reduced}
\end{equation} However, because we are interested in constraining a feature located in a low density region, for our main analysis we will assume the first-order approximation $g_i \simeq \gamma_i$ when fitting a model. From our source catalogues we extract the tangential component $g_{\text{t}}(\theta_j)$ in radial bins and estimate for each cluster the data covariance matrix as the sum of two terms: the first accounts for statistical noise in the average ellipticity and the second one takes into account the presence of additional shear introduced by uncorrelated structure along the line of sight. More details about the evaluation are presented in Appendix~\ref{app:noise}.

The top panel of Fig.~\ref{fig:shear} presents the average noise-weighted signal of the full cluster sample. The double error bars in the figure illustrate how the inclusion of the second source of noise has an impact on the uncertainties at large scales. An indicative NFW fit, obtained using the virial overdensity from \cite{Bryan1998} at an assumed redshift $z=0.25$, is also shown. The position of $r_{200m}$ for the best-fit model is also indicated in the same figure.

\subsection{Residual systematics}
\label{sec:sys}

In this section, we address the effects of the corrections we have implemented to tackle two systematic effects that are particularly important for our analysis: PSF anisotropy and cluster member contamination. In particular, we estimate the amplitude of any residual systematic effects as plotted in the bottom panel of Fig.~\ref{fig:shear}.  

In the KSB method, the observed galaxy polarizations are corrected for PSF anisotropy using
\begin{equation}
e_i \to e_i-  \sum_{j} P^\text{sm}_{ij} p_{j}^{\ast},
\end{equation}
where the smear polarizability $P^\text{sm}$ quantifies how susceptible a source is to PSF distortions and $p_j$ is the PSF anisotropy measured using point-like sources \citep[see, e.g., ][]{Hoekstra1998}. 

The observed polarizations and polarizabilities are, however, biased because of noise in the images. If unaccounted for, this leads to biased cluster masses. For the shear, these corrections can be expressed in terms of a multiplicative and additive bias, $\mu$ and $b$:
\begin{align}
\gamma_i  \to  (1 + \mu) \gamma_i  + b.
\end{align}

\begin{figure}
	\includegraphics[width=0.47\textwidth]{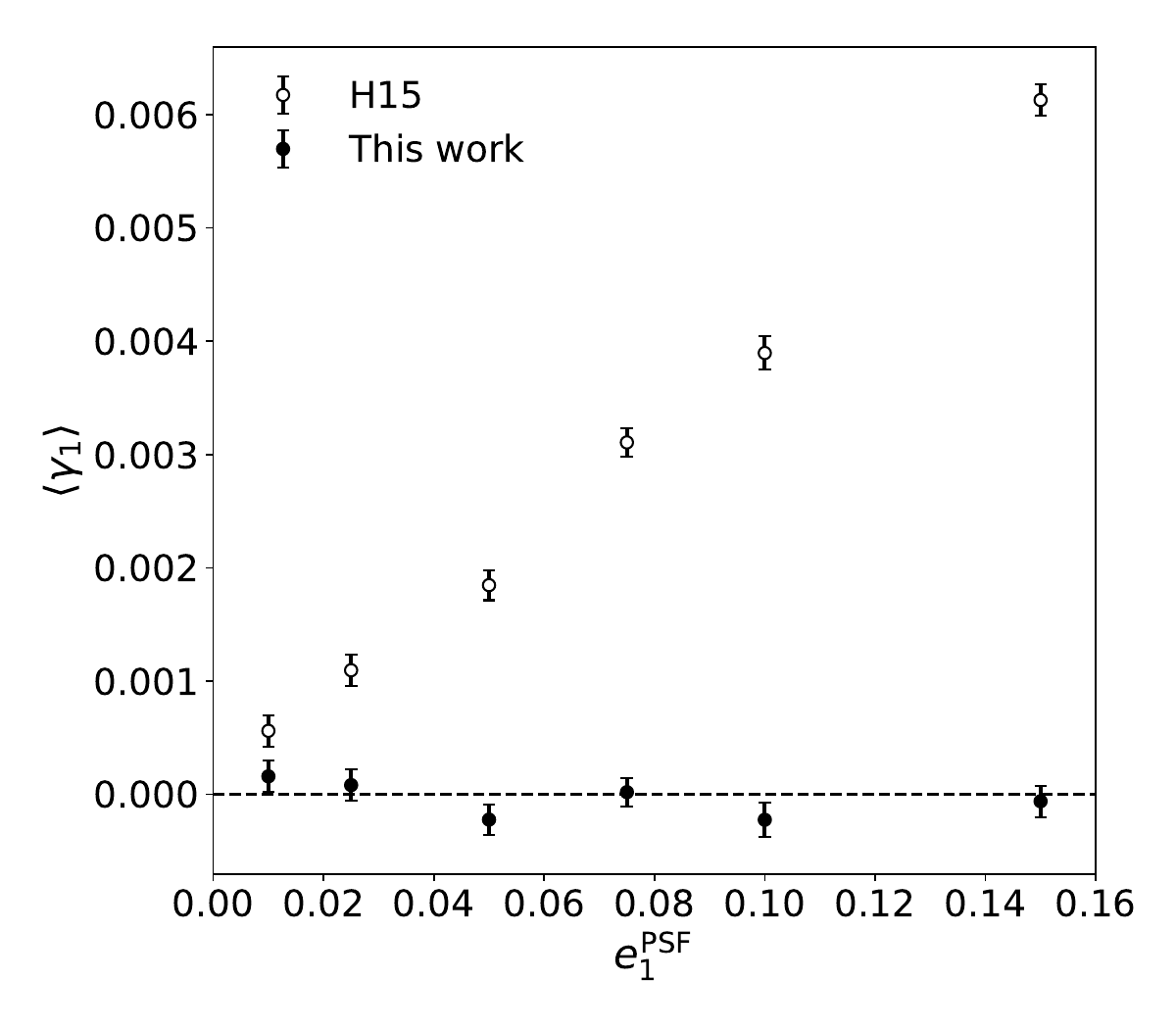}
	\caption{PSF correction improvements. Image simulations are used to quantify the residual additive bias not captured by the correction obtained in \citetalias{Hoekstra2015}. The circles show how residual additive bias in the average shear $\langle\gamma_1\rangle$ was present in the presence of simulated PSF anisotropy ($e^\text{PSF}_1\neq0$). In this work (filled points) we are able to nullify this effect by boosting the KSB smear polarizability $P^\text{sm}$. See Sec.~\ref{sec:sys} for more details.}
	\label{fig:changesb}
\end{figure}

To ensure accurate mass estimates, \citetalias{Hoekstra2015} focused on the impact of multiplicative bias. To do so, they 
used image simulations with a circular PSF to calibrate the bias as a function of source SNR and size. However, the actual PSF is not round and \citetalias{Hoekstra2015}, therefore, quantified the 
impact of an anisotropic PSF on the multiplicative bias correction. The details of these simulations, based on \texttt{galsim} \citep{Rowe2015}, can be found in section 2.2 and appendix~A of \citetalias{Hoekstra2015}. The galaxy properties are based on HST observations, resulting in  images that match the cluster data. The PSF is modelled as a Moffat profile, which is a good representation of ground based data. Appendix A in H15 examines the impact of PSF anisotropy and revealed that about 4 percent of this source of bias remains after correction (see their Fig. A1). While the impact of this residual bias is negligible, further study revealed that it can be reduced by empirically correcting the smear polarizability for noise bias. We have increased $P^\text{sm}$ by a factor of 1.065, such that no residual additive bias remains visible, see Fig.~\ref{fig:changesb}. We also verified that this latest correction does not introduce significant trends with source characteristics. We use the difference between the ensemble lensing signal measured before and after this improvement as a (conservative) estimate of any unknown systematics affecting the shape measurement method. 

The second effect we account for is the presence of cluster members in our source catalogues. Note that in this case, we have not updated the methodology from \citetalias{Hoekstra2015}, but we still report it here for completeness. If we assume that cluster members are randomly oriented, as found by \cite{Sifon2015}, their inclusion among our sources has the effect of diluting the measured shear. To correct for this, we multiply $\gamma_t(R)$ by a boost factor $B(R)$ defined as a function of the projected comoving distance $R$:
\begin{align}
B(R) = 1+ f_\text{cont}(R)/f_\text{obs}(R).
\end{align}
The contamination term $f_\text{cont}$ accounts for the decrease of the ellipticity average due to the presence unsheared sources and, by comparison with blank fields, it is found to be 1) a decreasing function of distance from the cluster centre and 2) negligible beyond a distance $r_\text{max}$. An extra factor $f_\text{obs}$ is also introduced to model the reduced background galaxy counts due to obscuration by cluster members. This factor is computed by stacking the cluster images with simulated blank fields and measuring how many simulated sources are obscured. 

The functions appearing in the boost factor are written empirically as:
\begin{align}
&\frac{1}{f_\text{obs}(R)} = 1+ \frac{0.021}{0.14 + (R/r_{500})^2}~\mathrm{and}
\\
&f_\text{cont}(R) = n_0 \left( \frac{1}{R+R_c} - \frac{1}{r_\text{max} + R_c}\right);
\end{align}
where $n_0$ and $R_c$ are fitted independently for each cluster and $B(R)=0$ for $R>r_\mathrm{max}\equiv 4(1+z)$ Mpc.

To quantify the amplitude of residual systematics for this second correction, we refer to \citetalias{Hoekstra2015}, where a residual scatter of about $2$ per cent around the ensemble correction was reported.

\section{Splashback}
\label{sec:splash}

\begin{figure*}
	\includegraphics[width=1.\textwidth]{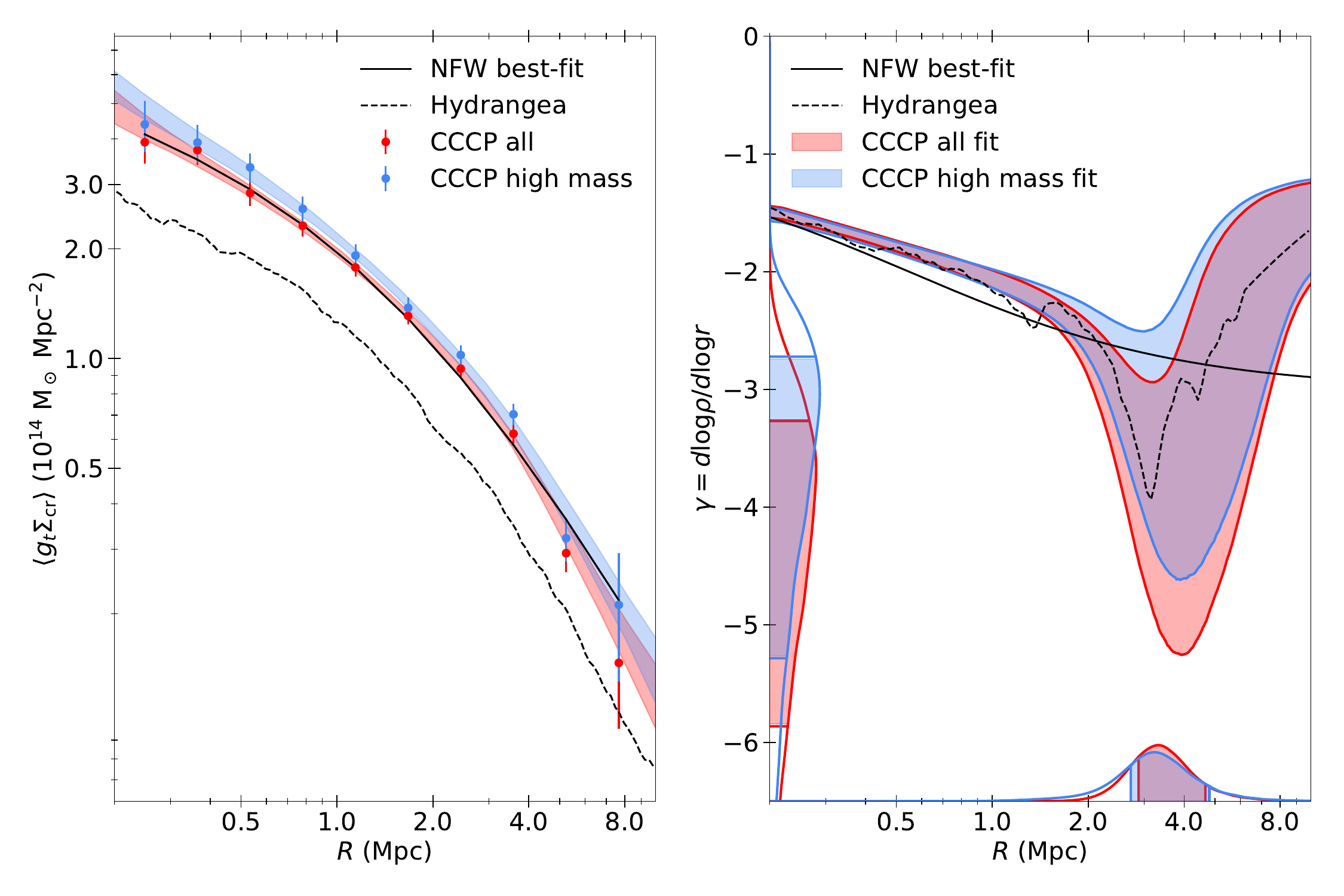}
	\caption{Splashback measurement. The left panel shows the measured lensing signal for our full sample and a subsample of the $13$ most massive clusters as a function of comoving clustercentric distance, together with the $68$ per cent confidence intervals from the \citetalias{Diemer2014} fit. The right panel shows the posterior of the three-dimensional logarithmic slope for the same model. The histograms on the horizontal axis are the distributions of the inferred position for the minimum of $\gamma$ (i.e. the splashback radius $r_\text{sp}$), while the histograms on the vertical axis are the distributions of $\gamma(r_\text{sp})$. The solid black lines refer to the NFW fit shown in Fig.~\ref{fig:shear}, while the dashed lines correspond to predictions from hydrodynamical simulations of massive clusters (Hydrangea). The amplitude of the Hydrangea and CCCP signals are different because we match the virial mass of our observed sample at $z\gtrsim0.2$ with simulated clusters at $z=0$.}
	\label{fig:result}
\end{figure*}

In this main part of the paper, we fit the observed weak lensing signal using the spherical density profile presented in \citetalias{Diemer2014}. This profile is designed to reproduce  the expected flattening of the density profile at large radii due to the presence of infalling material, as seen in numerical simulations.

\subsection{Fitting procedure}
\label{sec:fitting}
The projected surface density profile $\Sigma (R)$ for a spherical lens with matter density $\rho(r)$ is:
\begin{equation}
\Sigma (R) = 2\int_0^{\infty} dr' \, \rho\left(\sqrt{r'^2 + R^2}\right),
\end{equation}
where we limit the integration range of the line of sight variable $r'$ to $[0, 40]$ Mpc for our numerical calculations. We also verify that the chosen upper limit has no effect on our results by repeating the analysis with a wider range $[0, 80]$ Mpc. For cosmological overdensities, this profile can be connected to the lensing signal through eq.~\ref{eq:shear} and eq.~\ref{eq:reduced}.

In this section we use a model for $\rho(r)$ first introduced by \citetalias{Diemer2014} with the following components: an Einasto profile $\rho_{\text{Ein}}$ \citep{Einasto1965} to model the inner dark matter halo, a transition term $f_{\text{trans}}(r)$ to capture a steepening effect at the halo edge and a power-law $\rho_{\text{out}}(r)$ to model the distribution of infalling material in the outer regions. The mathematical expressions are the following:
\begin{equation}
\rho(r) = 
\rho_{\text{Ein}}(r) f_{\text{trans}} (r) + \rho_{\text{out}} (r);
\end{equation}
\begin{equation}
\rho_{\text{Ein}}(r)  = \rho_{\text{s}} \exp \left( -\frac{2}{\alpha}\left[\left( \frac{r}{r_{\text{s}}} \right)^{\alpha} - 1  \right] \right),
\end{equation}
\begin{equation}
f_{\text{trans}} (r) =  \left[ 1+ \left(\frac{r}{r_\text{t}}\right)^{\beta}\right]^{-\gamma/\beta},
\end{equation}
\begin{equation}
\rho_{\text{out}} = \rho_0 \left( \frac{r}{r_0}\right)^{-s_e}.
\end{equation}

In \citetalias{Diemer2014} the infalling term includes an offset corresponding to the average matter density, but this is not present in our fitting function because the tangential shear in eq.~\ref{eq:shear} is completely insensitive to it. 

In its general form, this model depends on a large number of parameters. In order to reduce its degrees of freedom we, therefore, choose to set strong priors on a few parameters. As done in \cite{Baxter2017} and \cite{Chang2018} we do not fit both $\rho_0$ and $r_0$, but choose to fix one of them, as they are degenerate. We impose Gaussian priors $\log(0.2)\pm 0.1$, $\log(6)\pm0.2$ and $\log(4)\pm0.2$ on the logarithms of the exponents $\log\alpha, \log \beta$ and $\log \gamma$ respectively. The loose prior on the Einasto shape parameter $\alpha$ is motivated by dark matter only simulations and its $1\sigma$ interval covers the expected scatter due to the redshift and mass distribution of our sample \citep{Gao2008, Dutton2014a}, while for the exponents in the transition term the stringent priors are centred on the values suggested by \citetalias{Diemer2014}. We also set a Gaussian prior on the truncation radius $r_t$, $4\pm2$, based on the same results. The location of the median is based on the $r_{200m}$ inferred from our NFW fit and the selected standard deviation covers the expected range due to the mass distribution of our sample. Finally, based on previous measurements, we also set a minimum value of $1$ for the outer slope $s_e$ and a physically motivated minimum value of $0$ for the density parameters $\rho_s$ and $\rho_0$.

A rescaling of the radial coordinate with an overdensity radius (e.g. $r_{200m}$) is often employed when fitting the profile described above. We also attempt to rescale our coordinates with either $r_{500c}$ or $r_{200m}$, but due to the uncertainties on the individual cluster profiles, no rescaling results in the splashback feature being constrained with higher precision. Despite this, we still attempt to remove the redshift dependence of the average matter density of the Universe by using comoving coordinates.

We follow \cite{Umetsu2016} and do not include a miscentering term in our tangential shear model. In general, a shift in position of the cluster centres reported in Table~\ref{tab:sample} would cause a smoothing of the lensing profile in the central region. An estimate of the area affected by such an effect can be obtained by considering the difference between two independent estimators of the halo centre: the position of the brightest cluster galaxy or the X-ray luminosity peak. Our sample is found to be well centred (see M13) with the root mean square of the offset between the two $\sigma_\mathrm{off}=33$ kpc. For the scales plotted in Fig.~\ref{fig:result} we therefore do not expect our data to be affected by miscentering. 

A fit the to input data $\gamma_t(R)$ with the covariance matrix defined in Sec.~\ref{sec:Data} is performed by sampling the posterior distribution of the parameters $[\rho_s, r_s, \log \alpha, r_t, \log \beta, \log \gamma, \rho_0, s_e]$ using the Markov Chain Monte Carlo ensemble sampler emcee\footnote{\url{https://emcee.readthedocs.io/}.} (\citealt{Foreman-Mackey2013}, based on \citealt{Goodman2010}).

\subsection{Interpretation}
\label{sec:int}

Figure~\ref{fig:result} visually presents our results. The left panel shows the best-fit model to the lensing signal, while the right panel shows the posterior distribution of the inferred profile. To better highlight the splashback feature we choose to focus on the dimensionless logarithmic slope $\gamma = d\log \rho/d\log r = r / \rho \; d \rho / dr$ when plotting the posterior of our model. 

For both CCCP samples considered a minimum of the slope is identified. At larger distances, the results are the least interesting. In these regions, the power-law term becomes dominant and the value of the slope is set exclusively by the exponent $s_e$. In particular, its lower limit is artificially imposed by our prior. 

What is more relevant to our study is the minimum value of the slope $\gamma(r)$ and its location, i.e., the splashback radius $r_\text{sp}$. The $68$ per cent credible intervals of both quantities are indicated as shaded sections of the vertical and horizontal histograms. Our measured $99.7$ per cent confidence interval of $\gamma(r_\text{sp})$ for the full sample is $[-10.9, -2.3]$, meaning that we are unable to measure a significant departure from the slope expected for an NFW profile (about $-2.5$). Despite this, we are still able to constrain the value of both the splashback radius and the logarithmic slope at this point, $r_\text{sp}=3.5^{+1.1}_{-0.7}$ Mpc and $\gamma(r_\text{sp}) = -4.3^{+1.0}_{-1.5}$. We also highlight that the high-mass sample returns similar constraints with only half the sample size, $r_\text{sp}=3.5^{+1.3}_{-0.8}$ and $\gamma(r_\text{sp}) = -3.7^{+0.9}_{-1.6}$. 

As a point of reference, we also show the expected profiles from a suite of zoom-in hydrodynamical simulations of massive clusters \citep[Hydrangea,][]{Bahe2017}. From the full Hydrangea sample, we have selected the $8$ most massive clusters for this comparison in order to obtain a sample with an average value of $\langle M_{200m} \rangle = 1.7\times10^{15}$ M$_\odot$, similar to our dataset, but evaluated at $z=0$ instead of $z=0.2$. Note that the amplitude of the signal plotted in Fig.~\ref{fig:result} is lower than the observed sample due the evolution of the average matter density of the Universe. Our slope measurements are found to be agreement with what is seen in simulations. 

As done in \cite{Umetsu2016}, we study the impact of the model parameters on the predictions for $r_\mathrm{sp}$ and $\gamma(r_\mathrm{sp})$ to verify that our dataset is informative and we are not simply sampling our model priors. Of crucial importance is the truncation radius $r_t$, which, in the original definition of the \citetalias{Diemer2014} profile, explicitly sets the position of the splashback feature. 

\begin{figure}
  \includegraphics[width=0.47\textwidth]{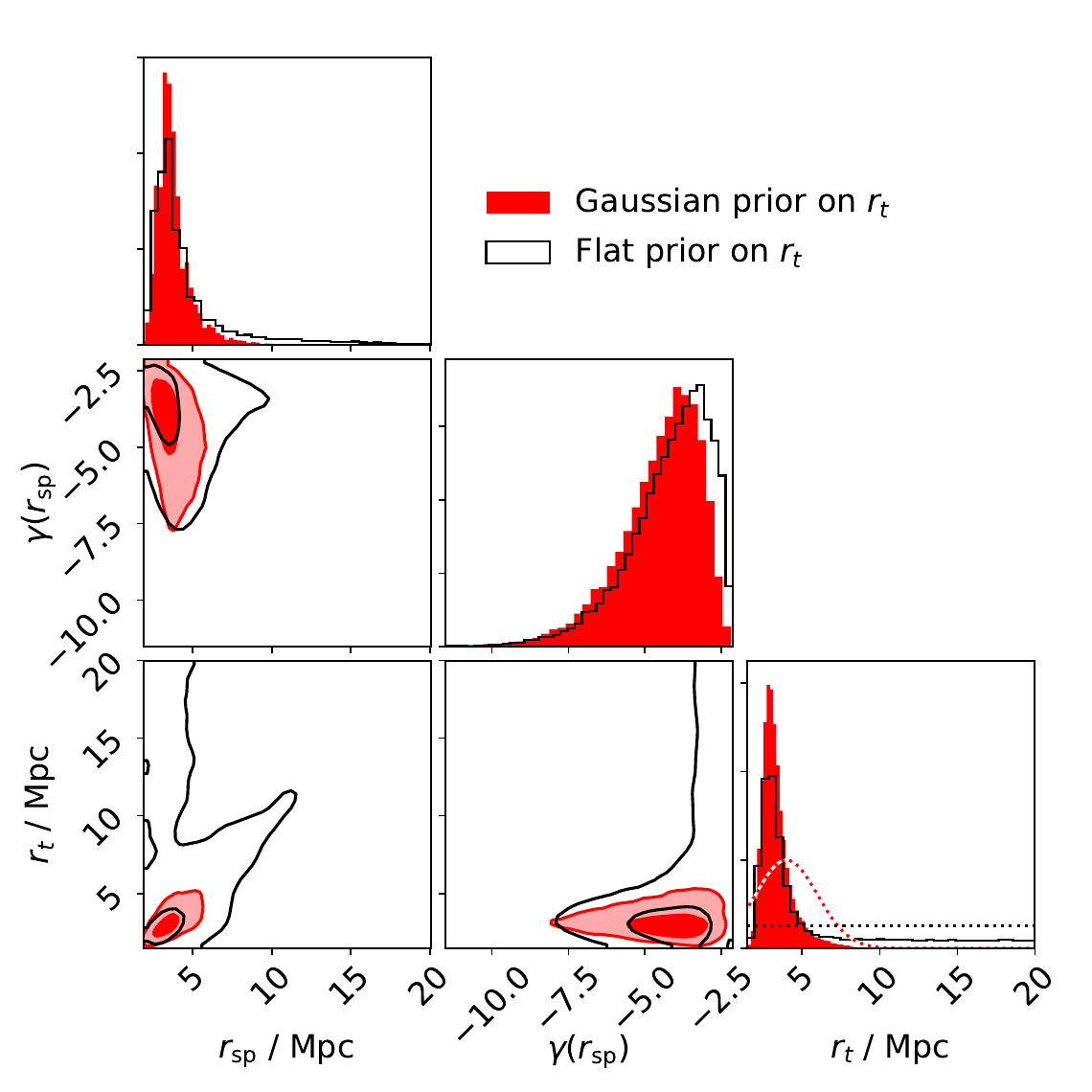}
  \caption{Impact of the prior on the truncation radius $r_\text{t}$ on our results. The corner plot presents the two-dimensional and marginalized posterior distributions for the \citetalias{Diemer2014} parameter $r_t$, the inferred splashback position $r_\mathrm{sp}$, and logarithmic slope $\gamma(r_\mathrm{sp})$. If, instead of a Gaussian prior (dashed red line), a flat prior is assumed (dashed black line), the parameter $r_t$ has no upper bound. This translates into weaker constraints on $r_\mathrm{sp}$.}
  \label{fig:post}
\end{figure}

Similarly to \cite{Umetsu2016}, we also find that we are unable to fully constrain this parameter. This can be seen in Fig.~\ref{fig:post}, where we plot the posteriors of three relevant parameters for two different choices of the $r_t$ prior: the Gaussian assumed in our main study and a flat prior in the range $[0, 20]$ Mpc. While the posterior for $\gamma(r_\mathrm{sp})$ (middle row) is mostly unaffected by this choice, we obtain a looser upper limit on the splashback radius (top panel) in the second case: $r_\text{sp} = 3.9^{+2.4}_{-0.9}$. As visible in the bottom-left panel, this is due to a clear correlation with $r_t$. 

We find no correlation between $r_\text{sp}$ and $r_{\text{t}}$ for $r_t\gtrsim10$ Mpc. In this regime, the location of the minimum of $\gamma(r)$ is controlled by the presence of the infalling term $\rho_{in}(r) \propto r^{-s_e}$. Because the slope $s_e$ is relatively gentle, if $r_t$ is large enough the truncation happens in a region dominated by the infalling material and cannot be constrained. Because the truncation is expected to be visible in the transition regime, our Gaussian prior on $r_t$ effectively forces it to a physically motivated position and, from the figure, we confirm that it does not introduce a biased posterior peak.

\section{Conclusions}
\label{sec:Conclusions}

We have shown in this work that targeted weak lensing observations of massive clusters can be used to measure the splashback feature and that particular care is required when correcting for residual PSF contaminations, which should be well understood, and estimating the data covariance matrix, which should take into account the presence of additional structure along the line of sight. Using a stack of $27$ massive clusters from CCCP we have fully constrained for the first time the splashback radius around massive clusters, $r_\text{sp}=3.6^{+1.2}_{-0.7}$, and similar precision has also been achieved with as little as $13$ objects. We stress that, because of the purely gravitational nature of weak lensing, minimal assumptions are required to interpret our signal.

In the last few years, the study of the physics of accretion at the outskirts of massive dark matter haloes has become observationally viable. Splashback offers a unique view into the phase-space configuration of haloes, which has not yet been explored in observations. In particular, the physics behind it appears to be remarkably uncomplicated and semi-analytical models of spherical collapse for cold dark matter are able to reproduce the expectations from N-body simulations \citep[e.g.,][]{Adhikari2014, Shi2016}. The fact that these results are based only on the dynamics of collapsing dark matter in an expanding Universe makes splashback a remarkable prediction of general relativity and dark matter. More generally, its connection to the growth of cosmological structures makes it a test for $\Lambda$CDM. As an example, it has also been shown recently that modifications of gravity have a significant impact on this feature \citep{Adhikari2018}. As the first results are starting to appear in the literature, we argue that splashback solicits further investigation exactly because it is a falsifiable prediction of the current paradigm.

We found that at the relevant scales a significant contribution to the lensing signal is cosmic noise. In the near future, this term can be reduced significantly with larger cluster samples. Looking further ahead, deep wide-area surveys such as {\it Euclid} \citep{Laureijs11} and LSST \citep{LSST} will provide unprecedented depth and survey area, and thus deliver the data required to study splashback over a wider mass and redshift range. 

\section*{Acknowledgements}

We thank Keiichi Umetsu, Andrej Dvornik and Koenraad Kuijken for useful discussion. OC is supported by a de Sitter Fellowship of the Netherlands Organization for Scientific Research (NWO). YMB acknowledges funding from the EU Horizon 2020 research and innovation programme under Marie Sk{\l}odowska-Curie grant agreement 747645 (ClusterGal) and the Netherlands Organisation for Scientific Research (NWO) through VENI grant 016.183.011. The Hydrangea simulations were in part performed on the German federal maximum performance computer ``HazelHen'' at the maximum performance computing centre Stuttgart (HLRS), under project GCS-HYDA / ID 44067 financed through the large-scale project ``Hydrangea'' of the Gauss Center for Supercomputing. Further simulations were performed at the Max Planck Computing and Data Facility in Garching, Germany. 

\appendix
\section{Noise covariance matrix}

For each cluster we model the noise covariance matrix for the lensing signal as the sum of two components:
\label{app:noise}
\begin{equation}
\mathbfss{C} = \mathbfss{C}^{\text{stat}} + \mathbfss{C}^{\text{lss}}.
\label{eq:cov}
\end{equation}

\begin{figure}
    \centering
    \includegraphics[width=0.5\textwidth]{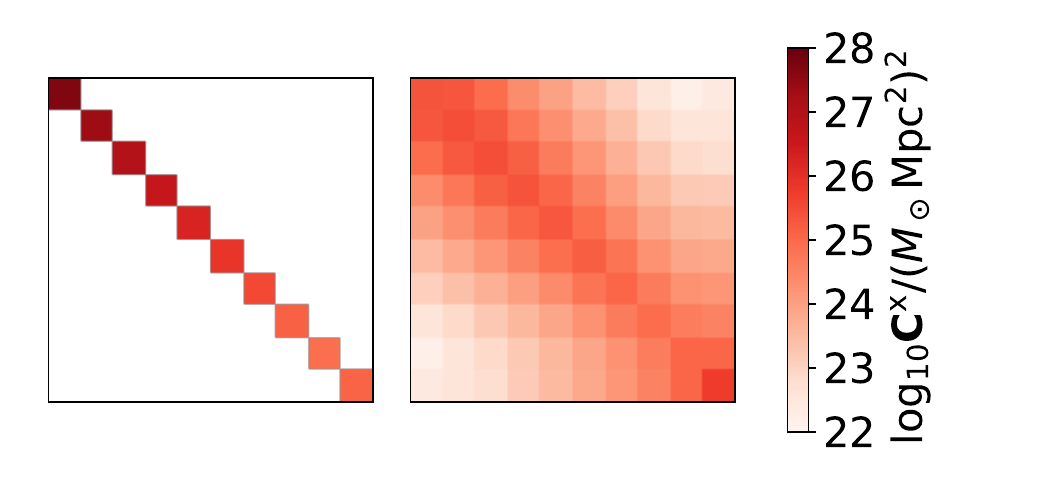}
    \caption{Covariance matrix. Visualization of the two components of the covariance matrix $\mathbfss{C} = \mathbfss{C}^\mathrm{stat} + \mathbfss{C}^\mathrm{lss}$ for the data points plotted in Fig.~\ref{fig:shear}. The diagonal matrix (left) is the statistical error $\mathbfss{C}^\mathrm{stat}$, the second one (right) is the  component due to uncorrelated structure along the line of sight, $\mathbfss{C}^\mathrm{lss}$. The top-left corner corresponds to the first data-point.}
    \label{fig:app}
\end{figure}

The first is a diagonal matrix accounting for the statistical error on the weighted average of the measured ellipticities and the second quantifies the additional shear variance caused by the presence of cosmic structure between viewer and source \citep{Hoekstra2003, Umetsu2011}
\begin{equation}
\mathbfss{C}^{\text{lss}}_{i, j} = 2\pi \int_0^\infty d\ell \; \ell P_{\kappa}(\ell ) g(\ell, \theta_i)
g(\ell, \theta_j),
\end{equation}
where $P_{\kappa}(\ell)$ represents the projected convergence power spectrum for the multipole number $\ell$. For an angular bin $\theta$ extending from $\theta_{-}$ to $\theta_{+}$, $g(l, \theta)$ is defined using the Bessel functions of the first kind of order zero and one, $J_0$ and $J_1$:

\begin{align}
\begin{split}
  g(\ell, \theta) =& 
  \left[ \frac{1-2\ln \theta_{-}}{\pi (\theta_{+}^2 - \theta_{-}^2)} \right]  \frac{\theta_{-} J_1(\ell \theta_{-})}{\ell}-
  \left[ \frac{1-2\ln \theta_{+}}{\pi (\theta_{+}^2 - \theta_{-}^2)} \right] 
   \frac{\theta_{+} J_1(\ell \theta_{+})}{\ell}
  \\
  -&
   \frac{2}{\pi (\theta_{+}^2 - \theta_{-}^2)}  
   \int_{\theta_1}^{\theta_2} d\phi \; \phi \log \phi J_0(l\phi).
 \end{split}
\end{align}
For a given cosmology, $P_{\kappa}(\ell)$ can be evaluated using the Limber projection starting from a source redshift distribution and a model for the non-linear matter power-spectrum \citep{Kilbinger15}. For this work, this is done using CAMB\footnote{\url{https://camb.info/}.} \citep{Lewis2013} and HALOFIT \citep{Takahashi2012}. As an example, the resulting covariance matrices for the average signal in Fig. 1 are presented in Fig.~\ref{fig:app}. 

A third term accounting for the intrinsic variance in a particular realization of galaxy clusters should be added to the matrix in Eq.~\ref{eq:cov}. For massive clusters in the considered redshift range, this term is found to be dominated by Poissonian scatter in the number of haloes contained within the correlated neighbourhood \citep{Gruen2015}. We neglect this term because in similar lensing analyses \citep[e.g.,][]{Umetsu2016a, Miyatake} it is always found to be sub-dominant to statistical and large-scale structure noise, especially on the scales of interest for this work.

\bibliographystyle{mnras}
\bibliography{splashback} 

\begin{thebibliography}{}
\makeatletter
\relax
\def\mn@urlcharsother{\let\do\@makeother \do\$\do\&\do\#\do\^\do\_\do\%\do\~}
\def\mn@doi{\begingroup\mn@urlcharsother \@ifnextchar [ {\mn@doi@}
  {\mn@doi@[]}}
\def\mn@doi@[#1]#2{\def\@tempa{#1}\ifx\@tempa\@empty \href
  {http://dx.doi.org/#2} {doi:#2}\else \href {http://dx.doi.org/#2} {#1}\fi
  \endgroup}
\def\mn@eprint#1#2{\mn@eprint@#1:#2::\@nil}
\def\mn@eprint@arXiv#1{\href {http://arxiv.org/abs/#1} {{\tt arXiv:#1}}}
\def\mn@eprint@dblp#1{\href {http://dblp.uni-trier.de/rec/bibtex/#1.xml}
  {dblp:#1}}
\def\mn@eprint@#1:#2:#3:#4\@nil{\def\@tempa {#1}\def\@tempb {#2}\def\@tempc
  {#3}\ifx \@tempc \@empty \let \@tempc \@tempb \let \@tempb \@tempa \fi \ifx
  \@tempb \@empty \def\@tempb {arXiv}\fi \@ifundefined
  {mn@eprint@\@tempb}{\@tempb:\@tempc}{\expandafter \expandafter \csname
  mn@eprint@\@tempb\endcsname \expandafter{\@tempc}}}

\bibitem[\protect\citeauthoryear{Ade et~al.,}{Ade et~al.}{2016}]{Ade2016}
Ade P. A.~R.,  et~al., 2016, \mn@doi [Astronomy {\&} Astrophysics]
  {10.1051/0004-6361/201525830}, 594, A13

\bibitem[\protect\citeauthoryear{Adhikari, Dalal  \& Chamberlain}{Adhikari
  et~al.}{2014}]{Adhikari2014}
Adhikari S.,  Dalal N.,   Chamberlain R.~T.,  2014, \mn@doi [Journal of
  Cosmology and Astroparticle Physics] {10.1088/1475-7516/2014/11/019}, 2014,
  019

\bibitem[\protect\citeauthoryear{Adhikari, Sakstein, Jain, Dalal  \&
  Li}{Adhikari et~al.}{2018}]{Adhikari2018}
Adhikari S.,  Sakstein J.,  Jain B.,  Dalal N.,   Li B.,  2018, Technical
  report, {Splashback in galaxy clusters as a probe of cosmic expansion and
  gravity}.
 (\mn@eprint {arXiv} {arXiv:1806.04302v1})

\bibitem[\protect\citeauthoryear{Bah{\'{e}} et~al.,}{Bah{\'{e}}
  et~al.}{2017}]{Bahe2017}
Bah{\'{e}} Y.~M.,  et~al., 2017, \mn@doi [Monthly Notices of the Royal
  Astronomical Society] {10.1093/mnras/stx1403}, 470, 4186

\bibitem[\protect\citeauthoryear{Baxter et~al.,}{Baxter
  et~al.}{2017}]{Baxter2017}
Baxter E.,  et~al., 2017, \mn@doi [The Astrophysical Journal]
  {10.3847/1538-4357/aa6ff0}, 841, 18

\bibitem[\protect\citeauthoryear{Bryan \& Norman}{Bryan \&
  Norman}{1998}]{Bryan1998}
Bryan G.~L.,  Norman M.~L.,  1998, \mn@doi [The Astrophysical Journal]
  {10.1086/305262}, 495, 80

\bibitem[\protect\citeauthoryear{Busch \& White}{Busch \&
  White}{2017}]{Busch2017}
Busch P.,  White S. D.~M.,  2017, \mn@doi [Monthly Notices of the Royal
  Astronomical Society] {10.1093/mnras/stx1584}, 470, 4767

\bibitem[\protect\citeauthoryear{{Chang} et~al.,}{{Chang}
  et~al.}{2018}]{Chang2018}
{Chang} C.,  et~al., 2018, \mn@doi [\apj] {10.3847/1538-4357/aad5e7}, \href
  {http://adsabs.harvard.edu/abs/2018ApJ...864...83C} {864, 83}

\bibitem[\protect\citeauthoryear{Cooray \& Sheth}{Cooray \&
  Sheth}{2002}]{Cooray2002}
Cooray A.,  Sheth R.,  2002, \mn@doi [Physics Reports]
  {10.1016/S0370-1573(02)00276-4}, 372, 1

\bibitem[\protect\citeauthoryear{{DES Collaboration} et~al.,}{{DES
  Collaboration} et~al.}{2017}]{DESCollaboration2017}
{DES Collaboration} D.,  et~al., 2017

\bibitem[\protect\citeauthoryear{Diemer \& Kravtsov}{Diemer \&
  Kravtsov}{2014}]{Diemer2014}
Diemer B.,  Kravtsov A.~V.,  2014, \mn@doi [The Astrophysical Journal]
  {10.1088/0004-637X/789/1/1}, 789, 1

\bibitem[\protect\citeauthoryear{Diemer, More  \& Kravtsov}{Diemer
  et~al.}{2013}]{Diemer2013}
Diemer B.,  More S.,   Kravtsov A.~V.,  2013, \mn@doi [The Astrophysical
  Journal] {10.1088/0004-637X/766/1/25}, 766, 25

\bibitem[\protect\citeauthoryear{Dutton \& Macci{\`{o}}}{Dutton \&
  Macci{\`{o}}}{2014}]{Dutton2014a}
Dutton A.~A.,  Macci{\`{o}} A.~V.,  2014, \mn@doi [Monthly Notices of the Royal
  Astronomical Society] {10.1093/mnras/stu742}, 441, 3359

\bibitem[\protect\citeauthoryear{Einasto}{Einasto}{1965}]{Einasto1965}
Einasto J.,  1965, Trudy Astrofizicheskogo Instituta Alma-Ata, 5, 87

\bibitem[\protect\citeauthoryear{Fillmore \& Goldreich}{Fillmore \&
  Goldreich}{1984}]{Fillmore1984}
Fillmore J.~A.,  Goldreich P.,  1984, \mn@doi [The Astrophysical Journal]
  {10.1086/162070}, 281, 1

\bibitem[\protect\citeauthoryear{Foreman-Mackey, Hogg, Lang  \&
  Goodman}{Foreman-Mackey et~al.}{2013}]{Foreman-Mackey2013}
Foreman-Mackey D.,  Hogg D.~W.,  Lang D.,   Goodman J.,  2013, \mn@doi
  [Publications of the Astronomical Society of the Pacific] {10.1086/670067},
  125, 306

\bibitem[\protect\citeauthoryear{Gao, Navarro, Cole, Frenk, White, Springel,
  Jenkins  \& Neto}{Gao et~al.}{2008}]{Gao2008}
Gao L.,  Navarro J.~F.,  Cole S.,  Frenk C.~S.,  White S. D.~M.,  Springel V.,
  Jenkins A.,   Neto A.~F.,  2008, \mn@doi [Monthly Notices of the Royal
  Astronomical Society] {10.1111/j.1365-2966.2008.13277.x}, 387, 536

\bibitem[\protect\citeauthoryear{Goodman \& Weare}{Goodman \&
  Weare}{2010}]{Goodman2010}
Goodman J.,  Weare J.,  2010, \mn@doi [Communications in Applied Mathematics
  and Computational Science] {10.2140/camcos.2010.5.65}, 5, 65

\bibitem[\protect\citeauthoryear{Gruen, Seitz, Becker, Friedrich  \&
  Mana}{Gruen et~al.}{2015}]{Gruen2015}
Gruen D.,  Seitz S.,  Becker M.~R.,  Friedrich O.,   Mana A.,  2015, \mn@doi
  [Monthly Notices of the Royal Astronomical Society] {10.1093/mnras/stv532},
  449, 4264

\bibitem[\protect\citeauthoryear{Gunn \& Gott}{Gunn \& Gott}{1972}]{Gunn1972}
Gunn J.~E.,  Gott J. R.~I.,  1972, \mn@doi [The Astrophysical Journal]
  {10.1086/151605}, 176, 1

\bibitem[\protect\citeauthoryear{Hoekstra}{Hoekstra}{2003}]{Hoekstra2003}
Hoekstra H.,  2003, \mn@doi [Monthly Notices of the Royal Astronomical Society]
  {10.1046/j.1365-8711.2003.06264.x}, 339, 1155

\bibitem[\protect\citeauthoryear{Hoekstra, Franx, Kuijken  \& Squires}{Hoekstra
  et~al.}{1998}]{Hoekstra1998}
Hoekstra H.,  Franx M.,  Kuijken K.,   Squires G.,  1998, \mn@doi [The
  Astrophysical Journal] {10.1086/306102}, 504, 636

\bibitem[\protect\citeauthoryear{Hoekstra, Franx  \& Kuijken}{Hoekstra
  et~al.}{2000}]{Hoekstra2000}
Hoekstra H.,  Franx M.,   Kuijken K.,  2000, \mn@doi [The Astrophysical
  Journal] {10.1086/308556}, 532, 88

\bibitem[\protect\citeauthoryear{Hoekstra, Mahdavi, Babul  \&
  Bildfell}{Hoekstra et~al.}{2012}]{Hoekstra2012}
Hoekstra H.,  Mahdavi A.,  Babul A.,   Bildfell C.,  2012, \mn@doi [Monthly
  Notices of the Royal Astronomical Society]
  {10.1111/j.1365-2966.2012.22072.x}, 427, 1298

\bibitem[\protect\citeauthoryear{Hoekstra, Bartelmann, Dahle, Israel, Limousin
  \& Meneghetti}{Hoekstra et~al.}{2013}]{Hoekstra2013a}
Hoekstra H.,  Bartelmann M.,  Dahle H.,  Israel H.,  Limousin M.,   Meneghetti
  M.,  2013, \mn@doi [Space Science Reviews] {10.1007/s11214-013-9978-5}, 177,
  75

\bibitem[\protect\citeauthoryear{Hoekstra, Herbonnet, Muzzin, Babul, Mahdavi,
  Viola  \& Cacciato}{Hoekstra et~al.}{2015}]{Hoekstra2015}
Hoekstra H.,  Herbonnet R.,  Muzzin A.,  Babul A.,  Mahdavi A.,  Viola M.,
  Cacciato M.,  2015, \mn@doi [Monthly Notices of the Royal Astronomical
  Society] {10.1093/mnras/stv275}, 449, 685

\bibitem[\protect\citeauthoryear{Kaiser, Squires  \& Broadhurst}{Kaiser
  et~al.}{1995}]{Kaiser1995}
Kaiser N.,  Squires G.,   Broadhurst T.,  1995, \mn@doi [The Astrophysical
  Journal] {10.1086/176071}, 449, 460

\bibitem[\protect\citeauthoryear{Kilbinger}{Kilbinger}{2015}]{Kilbinger15}
Kilbinger M.,  2015, \mn@doi [Reports on Progress in Physics]
  {10.1088/0034-4885/78/8/086901}, 78, 086901

\bibitem[\protect\citeauthoryear{{LSST Science Collaboration} et~al.,}{{LSST
  Science Collaboration} et~al.}{2009}]{LSST}
{LSST Science Collaboration} et~al., 2009, preprint, \href
  {http://adsabs.harvard.edu/abs/2009arXiv0912.0201L} {} (\mn@eprint {arXiv}
  {0912.0201})

\bibitem[\protect\citeauthoryear{Laigle et~al.,}{Laigle
  et~al.}{2016}]{Laigle2016}
Laigle C.,  et~al., 2016, \mn@doi [The Astrophysical Journal Supplement Series]
  {10.3847/0067-0049/224/2/24}, 224, 24

\bibitem[\protect\citeauthoryear{{Laureijs} et~al.,}{{Laureijs}
  et~al.}{2011}]{Laureijs11}
{Laureijs} R.,  et~al., 2011, preprint, \href
  {http://adsabs.harvard.edu/abs/2011arXiv1110.3193L} {} (\mn@eprint {arXiv}
  {1110.3193})

\bibitem[\protect\citeauthoryear{Lewis}{Lewis}{2013}]{Lewis2013}
Lewis A.,  2013, \mn@doi [Physical Review D] {10.1103/PhysRevD.87.103529}, 87,
  103529

\bibitem[\protect\citeauthoryear{Luppino \& Kaiser}{Luppino \&
  Kaiser}{1997}]{Luppino1997}
Luppino G.~A.,  Kaiser N.,  1997, \mn@doi [The Astrophysical Journal]
  {10.1086/303508}, 475, 20

\bibitem[\protect\citeauthoryear{Mahdavi, Hoekstra, Babul, Bildfell, Jeltema
  \& Henry}{Mahdavi et~al.}{2013}]{Mahdavi2013}
Mahdavi A.,  Hoekstra H.,  Babul A.,  Bildfell C.,  Jeltema T.,   Henry J.~P.,
  2013, \mn@doi [The Astrophysical Journal] {10.1088/0004-637X/767/2/116}, 767,
  116

\bibitem[\protect\citeauthoryear{Miyatake, More, Takada, Spergel, Mandelbaum,
  Rykoff  \& Rozo}{Miyatake et~al.}{2016}]{Miyatake2016}
Miyatake H.,  More S.,  Takada M.,  Spergel D.~N.,  Mandelbaum R.,  Rykoff
  E.~S.,   Rozo E.,  2016, \mn@doi [Physical Review Letters]
  {10.1103/PhysRevLett.116.041301}, 116, 041301

\bibitem[\protect\citeauthoryear{Miyatake et~al.,}{Miyatake
  et~al.}{2018}]{Miyatake}
Miyatake H.,  et~al., 2018

\bibitem[\protect\citeauthoryear{More, Diemer  \& Kravtsov}{More
  et~al.}{2015}]{More}
More S.,  Diemer B.,   Kravtsov A.~V.,  2015, \mn@doi [The Astrophysical
  Journal] {10.1088/0004-637X/810/1/36}, 810, 36

\bibitem[\protect\citeauthoryear{More et~al.,}{More et~al.}{2016}]{More2016}
More S.,  et~al., 2016, \mn@doi [The Astrophysical Journal]
  {10.3847/0004-637X/825/1/39}, 825, 39

\bibitem[\protect\citeauthoryear{Navarro, Frenk  \& White}{Navarro
  et~al.}{1997}]{Navarro1997}
Navarro J.~F.,  Frenk C.~S.,   White S. D.~M.,  1997, \mn@doi [The
  Astrophysical Journal] {10.1086/304888}, 490, 493

\bibitem[\protect\citeauthoryear{{Rowe} et~al.,}{{Rowe}
  et~al.}{2015}]{Rowe2015}
{Rowe} B.~T.~P.,  et~al., 2015, \mn@doi [Astronomy and Computing]
  {10.1016/j.ascom.2015.02.002}, \href
  {https://ui.adsabs.harvard.edu/\#abs/2015A&C....10..121R} {10, 121}

\bibitem[\protect\citeauthoryear{Rykoff et~al.,}{Rykoff
  et~al.}{2014}]{Rykoff2014}
Rykoff E.~S.,  et~al., 2014, \mn@doi [The Astrophysical Journal]
  {10.1088/0004-637X/785/2/104}, 785, 104

\bibitem[\protect\citeauthoryear{Shi}{Shi}{2016}]{Shi2016}
Shi X.,  2016, \mn@doi [Monthly Notices of the Royal Astronomical Society]
  {10.1093/mnras/stw925}, 459, 3711

\bibitem[\protect\citeauthoryear{Sif{\'{o}}n, Hoekstra, Cacciato, Viola,
  K{\"{o}}hlinger, van~der Burg, Sand  \& Graham}{Sif{\'{o}}n
  et~al.}{2015}]{Sifon2015}
Sif{\'{o}}n C.,  Hoekstra H.,  Cacciato M.,  Viola M.,  K{\"{o}}hlinger F.,
  van~der Burg R. F.~J.,  Sand D.~J.,   Graham M.~L.,  2015, \mn@doi [Astronomy
  {\&} Astrophysics] {10.1051/0004-6361/201424435}, 575, A48

\bibitem[\protect\citeauthoryear{Takahashi, Sato, Nishimichi, Taruya  \&
  Oguri}{Takahashi et~al.}{2012}]{Takahashi2012}
Takahashi R.,  Sato M.,  Nishimichi T.,  Taruya A.,   Oguri M.,  2012, \mn@doi
  [The Astrophysical Journal] {10.1088/0004-637X/761/2/152}, 761, 152

\bibitem[\protect\citeauthoryear{Umetsu \& Diemer}{Umetsu \&
  Diemer}{2017}]{Umetsu2016}
Umetsu K.,  Diemer B.,  2017, \mn@doi [The Astrophysical Journal]
  {10.3847/1538-4357/aa5c90}, 836, 231

\bibitem[\protect\citeauthoryear{Umetsu, Broadhurst, Zitrin, Medezinski, Coe
  \& Postman}{Umetsu et~al.}{2011}]{Umetsu2011}
Umetsu K.,  Broadhurst T.,  Zitrin A.,  Medezinski E.,  Coe D.,   Postman M.,
  2011, \mn@doi [The Astrophysical Journal] {10.1088/0004-637X/738/1/41}, 738,
  41

\bibitem[\protect\citeauthoryear{Umetsu, Zitrin, Gruen, Merten, Donahue  \&
  Postman}{Umetsu et~al.}{2016}]{Umetsu2016a}
Umetsu K.,  Zitrin A.,  Gruen D.,  Merten J.,  Donahue M.,   Postman M.,  2016,
  \mn@doi [The Astrophysical Journal] {10.3847/0004-637X/821/2/116}, 821, 116

\bibitem[\protect\citeauthoryear{Zu, Mandelbaum, Simet, Rozo  \& Rykoff}{Zu
  et~al.}{2017}]{Zu2017}
Zu Y.,  Mandelbaum R.,  Simet M.,  Rozo E.,   Rykoff E.~S.,  2017, \mn@doi
  [Monthly Notices of the Royal Astronomical Society] {10.1093/mnras/stx1264},
  470, 551

\bibitem[\protect\citeauthoryear{de Jong et~al.,}{de~Jong
  et~al.}{2017}]{DeJong2017}
de Jong J. T.~A.,  et~al., 2017, \mn@doi [Astronomy {\&} Astrophysics]
  {10.1051/0004-6361/201730747}, 604, A134

\makeatother
\end{thebibliography}


\bsp	
\label{lastpage}
\end{document}